# Glasslike Behavior in Aqueous Electrolyte Solutions

*David A. Turton,* [1] *Johannes Hunger,* [2] *Glenn Hefter,* [3] *Richard Buchner,* [2] *Klaas Wynne* [1]
[1] *Department of Physics, SUPA, University of Strathclyde, Glasgow G4 0NG, UK*
[2] *Institut für Physikalische und Theoretische Chemie, Universität Regensburg, D-93040 Regensburg, Germany*
[3] *Chemistry Department, Murdoch University, Murdoch, WA 6150, Australia*

When salts are added to water, the viscosity generally increases suggesting the ions increase the strength of the water's hydrogen-bond network. However, infrared pump-probe measurements on electrolyte solutions have found that ions have no influence on the rotational dynamics of water molecules implying no enhancement or breakdown of the hydrogen-bond network. Here we report optical Kerr-effect and dielectric relaxation spectroscopic measurements, which have enabled us to separate the effects of rotational and transitional motions of the water molecules. These data show that electrolyte solutions behave like a supercooled liquid approaching a glass transition in which rotational and translational molecular motions are decoupled. It is now possible to understand previously conflicting viscosity data, nuclear magnetic resonance relaxation, and ultrafast infrared spectroscopy in a single unified picture.

It is well known that when salts are added to water the viscosity typically increases, suggesting that the ions alter the hydrogen-bond network of the water,[1] which appears to be confirmed by, for example, neutron diffraction experiments.[2] However, recent ultrafast infrared pump-probe measurements on electrolyte solutions have found that ions do not influence the rotational dynamics of water molecules suggesting that there is no enhancement or breakdown of the hydrogen-bond network in liquid water.[3] As the effect of ions and ionic moieties on the structure of water is important for understanding protein stability, enzymatic reactions, and substrate binding, it is crucial to resolve this paradox.[4] Here we report ultrafast optical Kerr effect[5] (OKE) and dielectric relaxation[6] (DR) spectroscopy measurements, which show that salt solutions behave like a supercooled liquid approaching a glass transition, where rotational and translational molecular motions become decoupled. The rotational motions of bulk water molecules – observed as an α-relaxation in DR – are essentially independent of concentration. The translational motions seen in OKE spectroscopy can be understood as a β-relaxation[7] and their dynamics become increasingly inhomogeneous with increasing salt concentration. This insight reconciles previously conflicting viscosity data,[8] nuclear magnetic resonance relaxation,[9] and ultrafast infrared spectroscopy[3] data in a single unifying picture.

When simple inorganic salts are added to water, the viscosity typically increases (see FIG. 1). In the relatively low concentration range (up to ~0.5 M), this is described by the semi-empirical Jones-Dole equation,[1] $\eta/\eta_0 = 1 + Ac^{1/2} + Bc$, which expresses the shear viscosity $\eta$ in terms of the viscosity of pure water $\eta_0$, salt concentration $c$, and parameters $A$ and $B$. For many salts additional terms have to be added to describe the rapid increase in viscosity at higher concentrations. The obvious conclusion to draw from this behavior is that ions alter the structure of water. Indeed, the empirical Jones-Dole $B$ coefficient is often used to classify ions as either structure makers (kosmotropes) or structure breakers (chaotropes).

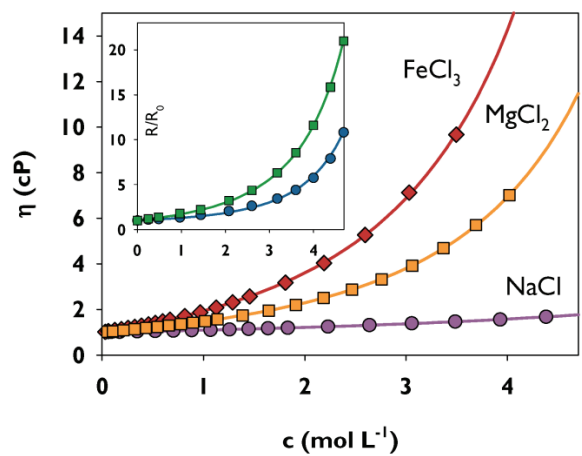

FIG. 1. (Color online) Dependence of viscosity and NMR relaxation rate on salt concentration. Points: viscosities[8] of aqueous solutions of NaCl (●), MgCl$_2$ (■), and FeCl$_3$ (◆). Lines: fits to Eq (2) with $c_0$/M = 12.7 (NaCl), 12.0 (MgCl$_2$), and 11.9 (FeCl$_3$) and Q = 0.6, 3.4, and 3.9 respectively. (**Inset**) Points: NMR quadrupolar relaxation rates of $^{25}$Mg$^{2+}$ (●) and $^{35}$Cl$^-$ (■) in aqueous MgCl$_2$ solutions normalized to the rate extrapolated to pure water.[9] Lines: fits to Eq (2) with $c_0$ = 12 M (fixed) and the Q = 4.5 (Mg$^{2+}$) and 4.1 (Cl$^-$).

The Stokes-Einstein-Debye equation $t_2 = V\eta/k_BT$ relates the macroscopic shear viscosity to the molecular volume $V$ and the diffusive rotational relaxation time $t_2$ as measured in a four-wave mixing spectroscopic experiment.[6] This predicts that as the viscosity increases with salt concentration, the rotational relaxation time would lengthen proportionally. However, ultrafast infrared pump-probe studies and their theoretical analysis have come to the surprising conclusion that the diffusive orientational relaxation of water outside the first solvation shell of the ions is **hardly affected** by the presence of salts.[3,10] This has been shown in aqueous solutions of Mg(ClO$_4$)$_2$, NaClO$_4$, and Na$_2$SO$_4$.[3] In all of these cases, infrared pump-probe experiments measure a 2.5 ps decay time for the orientational relaxation of water molecules in bulk water as distinct from water molecules in the relatively stable hydration shells surrounding the cations. Nuclear magnetic resonance spectroscopy (NMR) has been used[9] to measure the quadrupolar relaxation rates of $^{25}$Mg$^{2+}$ and $^{35}$Cl$^-$ ions in aqueous MgCl$_2$ solutions, which were found to increase with concentration approximately **in line with the increase in viscosity** (see the inset of FIG. 1). Thus, two different but widely used and reliable experimental techniques come to diametrically opposite conclusions: one does and one does not follow viscosity.

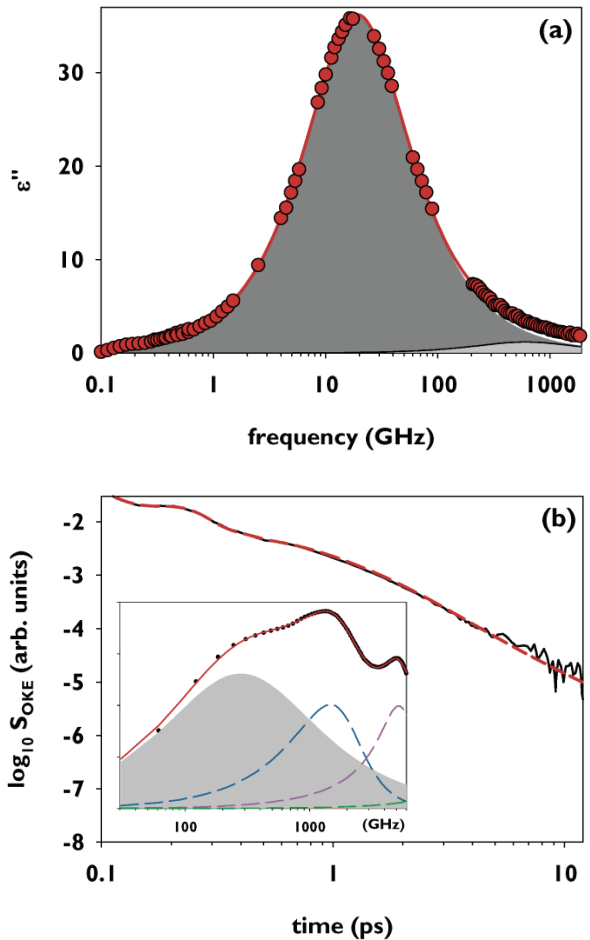

FIG. 2. (Color online) Dielectric relaxation (DR) and optical Kerr-effect (OKE) spectroscopy data on water at 25° C and their fits. (**a**) Dielectric loss spectrum, $\varepsilon''(\nu)$, of water fitted to the sum of two Debye equations with relaxation times of 8.38 ps (dark grey) and 0.3 ps (light grey).[11] (**b**) Relaxation of water measured by time-domain OKE spectroscopy (solid line) and fit (dashed line) by the Cole-Cole function and two damped harmonic-oscillator functions centered at 66 and 197 cm$^{-1}$. (**inset**) The OKE data transformed to the frequency domain: data (dots), fit (solid line), and a decomposition of the fit into Cole-Cole (grey area) and harmonic-oscillator functions (dashed lines).

This contradiction is resolved here by using ultrafast heterodyne-detected optical Kerr-effect[5] (OKE) and dielectric relaxation[6] (DR) spectroscopy to study aqueous solutions of NaCl and MgCl$_2$. OKE spectroscopy measures the two-point time-correlation function of the anisotropic part of the polarizability tensor in the time domain, while DR spectroscopy measures the two-point correlation function of the dipole-moment vector in the frequency domain.[12] In principle, both techniques measure the same dynamics – molecular reorientations – but with different amplitudes.[13]

The experimental setup for the OKE measurements has been described previously.[5,13] For this experiment, 800-nm 24-fs (FWHM) sech$^2$ pulses with 8 nJ per pulse at a repetition rate of 76 MHz were used. The beam was split into pump and probe beams (9:1), which were co-focused by a f=10-cm achromat into the sample contained in a 2-mm-pathlength quartz cuvette. The variable pump-probe time delay was introduced by an optical delay line with a resolution of 500 nm (3.3 fs). For each scan, 10 ps of baseline signal was measured before time-zero in order to obtain an accurate estimate of the baseline. Both pump and probe beams were mechanically chopped at rates of about 5 kHz in the ratio of 5:7 with lock-in demodulation at the difference frequency.

The dielectric spectra of aqueous MgCl$_2$ at 25 °C were obtained using a vector network analyzer (VNA) with a frequency range of 0.2-20 GHz[14] and two waveguide interferometers (IFMs) with a frequency range of 27-39 and 60-89 GHz.[15] The VNA was calibrated with air/mercury/water (open/short/load) and the raw data corrected with a Padé calibration using water, propylene carbonate, N,N-dimethylacetamide, and benzonitrile as secondary standards.[16] Samples were obtained by diluting a stock solution of $c$ = 4.398 mol/L. The latter was prepared from MgCl$_2$·6H$_2$O (Sigma Aldrich analytical grade) and Millipore water and its concentration determined by titration with a standard EDTA solution. Due to the high conductivity, reasonable VNA data could only be obtained above 0.6 GHz and the spectra show the typical systematic errors ("wiggles") for highly conducting samples.

FIG. 2(a) shows the DR spectrum of room-temperature water. The spectrum can be fit by a single Debye function $(1 - i\omega t_1)^{-1}$ with $t_1$ = 8.38 ps up to ~100 GHz where a weak secondary relaxation with $t_1'$ = 0.3 ps appears.[11,17] FIG. 2(b) shows the time-domain OKE decay in room-temperature water. This has previously been fit by different models including the Kohlrausch-Williams-Watts function,[18] but our data are most satisfactorily fit by a single Cole-Cole function

$$\left[1 - (i\omega t_2)^\beta\right]^{-1} \qquad (1)$$

with $t_2$ = 0.61 ps and $\beta$ = 0.86.[19]

The measured relaxation times $t_1$ and $t_2$ clearly differ. As DR spectroscopy measures a correlation function of vectors while OKE spectroscopy measures one of tensors, the relationship $t_2 = t_1/3$ is expected if the measured dynamics originate in single-molecule rotations. However, as we have shown elsewhere,[6] DR and OKE (or Raman) spectroscopy do not measure the same dynamics. The water molecule has a large dipole moment but an almost isotropic polarizability tensor.[20] Therefore, DR spectroscopy is sensitive to diffusive **orientational** relaxation of water molecules – often referred to as an α relaxation – giving a timescale of $t_1/3$ = 2.8 ps. Because of the near-isotropic molecular polarizability tensor, OKE is insensitive to pure rotational (single molecule) motions and instead measures only interaction-induced effects due to **translational** motions of pairs and larger groups of water molecules.[12,19] The good fit of the Cole-Cole function to the OKE data for water with $t_2$ = 0.61 ps (in between rotational diffusion at low frequency and librational motions at high frequency) shows that OKE spectroscopy measures a β relaxation related to the formation of transient cages in the liquid.[7,18] The Cole-Cole expo-nent $\beta$ indicates the degree of heterogeneity in the liquid. The complementarity of DR and OKE is the crucial aspect of this study. It holds with (near) perfection for water and to a much lesser extent for methanol.[6]

OKE and DR data were collected on a series of aqueous NaCl and $MgCl_2$ solutions at concentrations up to 4.5 M. The Cole-Cole function Eq (1) could be fit (in the time domain) to the OKE datasets at all concentrations. The concentration-dependent fit parameters are shown in FIG. 3(a). The $t_2$ relaxation time is seen to rise monotonically with concentration, becoming five times larger in 4.5 M $MgCl_2$ solution than in pure water. The Cole-Cole $\beta$ exponent decreases monotonically with concentration showing that the environment of the water molecules becomes increasingly inhomogeneous. These results are broadly consistent with spontaneous Raman scattering experiments.[21]

The DR data for aqueous solutions of $MgCl_2$ are well fit with the Cole-Cole function and also show increasing inhomogeneity with concentration (see FIG. 3(b)). Compared with $t_2$, the relaxation time $t_1/3$ shows surprisingly little variation with concentration. The slight decrease in $t_1/3$ above ~0.5 M is typical for aqueous solutions of small inorganic ions.[11] The subsequent slight increase in $t_1/3$ above ~3 M is such that it remains larger than $t_2$ at all concentrations. Thus, $t_1/3$ is essentially decoupled from the solution viscosity (FIG. 1).

The inset of FIG. 3(b) shows that the static dielectric constant decreases by a factor of 2.9 as a function of $MgCl_2$ concentration even though in 4 M $MgCl_2$ the concentration of water has only decreased to 49.6 M from the bulk value of 55.4 M.[8] This reduction is caused by the effective "removal" (immobilization) of water molecules from the bulk due to cation solvation.[11] The measured reduction in dielectric constant at 4.4 M corresponds with 7.1 water molecules being removed per cation. This broadly agrees with the coordination of the $Mg^{2+}$ ion by about 6 water molecules.[22,23] No DR signal arises from the $[Mg(H_2O)_6]^{2+}$ complex because the symmetric coordination causes a vanishing net dipole moment. On the other hand, DR studies have shown that the dynamics of water molecules in the coordination shells of most inorganic anions is virtually indistinguishable from bulk water dynamics.[11] The inset of FIG. 3(b) also shows the total concentration of water bound to the cations (calculated from the dielectric constant), which is seen to rise rapidly at low cation concentrations before saturating.

If the OKE signal originated only in the water molecules not bound to the cations, the amplitude of the β relaxation would decrease by a factor of three. In fact, the measured instantaneous hyperpolarizability increases by a factor of 1.7 consistent with the greater polarizability of chloride ions compared to water[8] and the amplitude of the β-relaxation component increases to a similar degree. This shows that, in contrast to DR spectroscopy, all water molecules (those bound to the cations as well as those in the bulk and near anions) contribute to the signal measured by OKE spectroscopy.

As can be seen in FIG. 3(a), the rate of change of $t_2$ increases with concentration with a rapid acceleration at high concentration. This behavior cannot be explained simply by the increase in bound water because – as the inset of FIG. 3(b) shows – the rate of change of the concentration of bound vs. bulk water decreases with concentration.

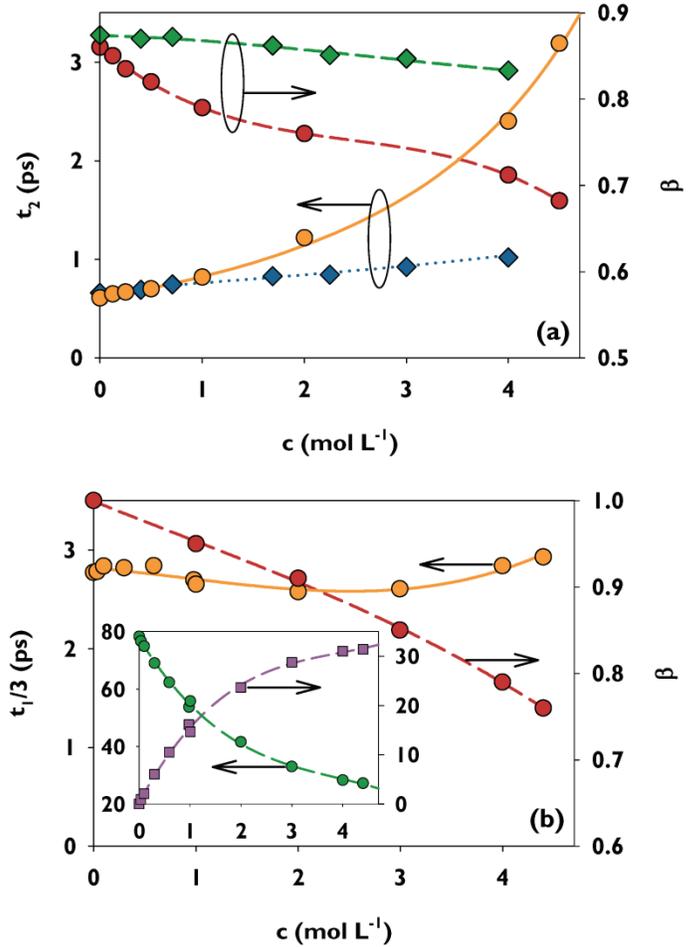

FIG. 3. (Color online) Fit parameters (using the Cole-Cole function, Eq (1)) for OKE and DR data as a function of salt concentration. **(a)** Fit parameters ($t_2$ and $\beta$) for OKE data on aqueous NaCl (◆) and $MgCl_2$ (●) solutions. The values of the relaxation-time $t_2$ have been fit to Eq (2) with $c_0$ fixed at 12.7 M (NaCl, dotted curve) or 12 M ($MgCl_2$, solid curve) and Q found as 0.07 (NaCl) and 2.0 ($MgCl_2$). **(b)** Fit parameters ($t_1/3$ and $\beta$) for DR data in aqueous $MgCl_2$ solutions. The dashed lines are in all cases added only as a guide to the eye. **(Inset)** The static dielectric constant measured with DR spectroscopy (●) and the calculated[11] concentration (mol $L^{-1}$) of water bound to the $Mg^{2+}$ cation (■) as functions of concentration.

Pure water has a glass transition temperature of ~135 K (obscured by a change of structure at 225 K), which increases with the addition of salts and other solutes.[24] For example, in ~5 M $MgCl_2$ solution, the glass-transition temperature is 170 K.[25] Measurements of electrical conductivity and viscosity as a function of temperature in aqueous electrolyte solutions have shown temperature-dependent behavior consistent with the empirical Vogel-Fulcher-Tammann (VFT) equation,[26] which predicts a critical glass-transition temperature. For viscosities under isothermal conditions, this can be expressed as a function of electrolyte concentration as[27]

$$\eta/\eta_0 = 1 + P\left\{e^{Qc_0/(c_0-c)} - e^Q\right\} \quad (2)$$

where $c_0$ is a glass-transition concentration and $P$ and $Q$ are empirical parameters.

The solid lines in FIG. 1 are fits of Eq (2) to the concentration-dependent viscosity. The increase in viscosity becomes more extreme in the series $Na^+$, $Mg^{2+}$, to $Fe^{3+}$, which may well be related to the residence time of water in the first hydration shell of the cation. Molecular dynamics computer simulations have estimated residence times of 14 ps [22] to 34 ps [28] for $Na^+$ and 228 ps [23] to 422 ps [22] for $Mg^{2+}$. The experimental residence time for $Fe^{3+}$ is ~$10^{-5}$–$10^{-3}$ s. [29] The fits to Eq (2) produce glass-transition concentrations $c_0$ of 12.7, 12.0, and 11.9 M for NaCl, $MgCl_2$, and $FeCl_3$ respectively, indicating the increased tendency for glass formation. The fits also give $Q$-parameters of 0.6, 3.4, and 3.9 respectively, consistent with an increasingly rapid approach to the glassy state. The NMR quadrupolar relaxation rates of $^{25}Mg^{2+}$ and $^{35}Cl^-$ ions in aqueous solution were also fit with Eq (2), as shown by the solid lines in the inset of FIG. 1, with an identical glass-transition concentration $c_0$ of 12.0 M. The $Q$-parameter was found to be somewhat larger at 4.5 ($Mg^{2+}$) and 4.1 ($Cl^-$), which could be due to ion-ion interactions at the highest concentrations. [9] Thus, it is reasonable and consistent to describe concentrated electrolyte solutions in terms of liquids close to a glass transition.

FIG. 3(a) shows Eq (2) fit to the $t_2$ relaxation-time parameters measured with OKE spectroscopy. Again, this is consistent with a glass-transition concentration $c_0$ of 12.0 M, although the $Q$-parameter was found to be smaller at 2.0. Thus, we find here that the β relaxation measured using OKE follows the same trend as the macroscopic shear viscosity and NMR quadrupolar relaxation. Crucially, the three measurements are consistent with an essentially identical glass-transition concentration. As OKE spectroscopy in water is sensitive to translational rather than rotational motions, these results are consistent with a slowing down of the translations of all water molecules in the electrolyte solution, and a complete arrest of translational motion at the glass-transition concentration.

In contrast, although DR in pure water follows the VFT equation over the temperature range of 0°-60° C (with a critical temperature of 131 K [30] close to the glass-transition temperature of water [24]), the expected slowing down with increasing salt concentration is not seen. These observations are therefore consistent with the picture of a "jamming" transition as observed for granular materials and colloidal suspensions. [31] Water molecules make and break hydrogen bonds with chloride ions on about the same timescale of ~3 ps as with each other, [10] whereas they remain in the first hydration shell of $Mg^{2+}$ for hundreds of picoseconds. [22,23] At a concentration of 4.4 M $MgCl_2$, there are overall ~11 water molecules per cation, [8] ~7 of which form a stable hydration shell around the cation (Fig. 3(b) inset) in which rotation is strongly impeded. The remaining 4 water molecules can be assumed to form small "pools" where rotation is still as free as in bulk water but translations are slowed down. At the glass-transition concentration, the density of clusters is so great that they jam, effectively turning the electrolyte solution into a glass. [31] This picture is consistent with the "colloidal suspension" picture proposed previously. [3]

This picture is also consistent with NMR measurements. [9] NMR quadrupolar relaxation originates in fluctuating electric-field gradients and the relaxation rate can be described by $R \propto \chi^2 \tau_c$, where $\chi$ is the strength and $\tau_c$ the correlation time of the electric-field-gradient fluctuations. [32] The quadrupolar relaxation of simple ions such as $Mg^{2+}$ and $Cl^-$ is in principle sensitive to all fluctuations of the dipole moments of local water molecules: rotations and translations. However, because the NMR lineshape is motionally narrowed, the slowest fluctuations have the greatest effect on the quadrupolar relaxation rate. Therefore, the NMR relaxation rate correlates with the slowing translational motions and thus increases with viscosity while the orientational relaxations remain fast. Thus, the glass model of electrolyte solutions shows that all the various measurements are in fact consistent.

## ACKNOWLEDGMENT

We gratefully acknowledge funding for this project from the Engineering and Physical Sciences Research Council (EPSRC).

1. H. D. B. Jenkins and Y. Marcus, Chem. Rev. **95** (8), 2695 (1995).
2. R. Mancinelli, A. Botti, F. Bruni, M. A. Ricci, and A. K. Soper, Phys. Chem. Chem. Phys. **9** (23), 2959 (2007).
3. A. W. Omta, M. F. Kropman, S. Woutersen, and H. J. Bakker, Science **301**, 347 (2003).
4. W. H. Qiu, Y. T. Kao, L. Y. Zhang, Y. Yang, L. J. Wang, W. E. Stites, D. P. Zhong, and A. H. Zewail, Proc. Natl. Acad. Sci. U. S. A. **103** (38), 13979 (2006).
5. G. Giraud, J. Karolin, and K. Wynne, Biophys. J. **85** (3), 1903 (2003).
6. T. Fukasawa, T. Sato, J. Watanabe, Y. Hama, W. Kunz, and R. Buchner, Phys. Rev. Lett. **95** (19), 197802 (2005).
7. P. G. Debenedetti and F. H. Stillinger, Nature **410** (6825), 259 (2001).
8. *Handbook of Chemistry and Physics*. (CRC Taylor and Francis, 2006).
9. R. Struis, J. Debleijser, and J. C. Leyte, J. Phys. Chem. **93** (23), 7943 (1989).
10. D. Laage and J. T. Hynes, Proc. Natl. Acad. Sci. **104** (27), 11167 (2007).
11. W. Wachter, W. Kunz, R. Buchner, and G. Hefter, J. Phys. Chem. A **109** (39), 8675 (2005).
12. C. J. Fecko, J. D. Eaves, and A. Tokmakoff, J. Chem. Phys. **117** (3), 1139 (2002).
13. G. Giraud and K. Wynne, J. Chem. Phys. **119** (22), 11753 (2003).
14. R. Buchner, G. T. Hefter, and P. M. May, J. Phys. Chem. A **103** (1), 1 (1999).
15. J. Barthel, K. Bachhuber, R. Buchner, H. Hetzenauer, and M. Kleebauer, Ber. Bunsen-Ges. Phys. Chem. Chem. Phys. **95** (8), 853 (1991).
16. S. Schrödle, G. Hefter, W. Kunz, and R. Buchner, Langmuir **22** (3), 924 (2006).
17. R. Buchner, J. Barthel, and J. Stauber, Chem. Phys. Lett. **306** (1-2), 57 (1999).
18. R. Torre, P. Bartolini, and R. Righini, Nature **428** (6980), 296 (2004).
19. D. A. Turton and K. Wynne, J. Chem. Phys. **in press** (2008).
20. W. F. Murphy, J. Chem. Phys. **67** (12), 5877 (1977).
21. Y. Wang and Y. Tominaga, J. Chem. Phys. **101** (5), 3453 (1994).
22. S. Obst and H. Bradaczek, J. Phys. Chem. **100** (39), 15677 (1996).
23. D. Jiao, C. King, A. Grossfield, T. A. Darden, and P. Y. Ren, J. Phys. Chem. B **110** (37), 18553 (2006).
24. C. A. Angell, Chem. Rev. **102** (8), 2627 (2002).
25. C. A. Angell and E. I. Sare, J. Chem. Phys. **52** (3), 1058 (1970).
26. C. A. Angell, K. L. Ngai, G. B. Mckenna, P. F. Mcmillan, and S. W. Martin, J. Appl. Phys. **88** (6), 3113 (2000).
27. C. A. Angell and R. D. Bressel, J. Phys. Chem. **76** (22), 3244 (1972).
28. K. B. Moller, R. Rey, M. Masia, and J. T. Hynes, J. Chem. Phys. **122** (11), 114508 (2005).
29. G. W. Neilson and J. E. Enderby, Adv. Inorg. Chem. **34**, 195 (1989).
30. W. J. Ellison, J. Phys. Chem. Ref. Data **36** (1), 1 (2007).
31. P. N. Segre, V. Prasad, A. B. Schofield, and D. A. Weitz, Phys. Rev. Lett. **86** (26), 6042 (2001).
32. A. Bagno, F. Rastrelli, and G. Saielli, Prog. Nucl. Mag. Res. Spectrosc. **47** (1-2), 41 (2005).